\documentclass{PoS}

\newcommand{\be}{\begin{eqnarray}}
\newcommand{\ee}{\end{eqnarray}}

\usepackage{amssymb,amsmath,cite}

\title{Aspects of the new Fayet--Iliopoulos terms} 

\ShortTitle{New FI terms}

\author{\speaker{Fotis Farakos}
\\ KU Leuven, Institute for Theoretical Physics, \\
			Celestijnenlaan 200D, B-3001 Leuven, Belgium 
        \\ 
        E-mail: \email{fotisfgm@gmail.com}}

\abstract{We review the properties of the new type of Fayet--Iliopoulos terms that do not require the gauging of the R-symmetry when embedded in N=1 supergravity, 
and we discuss recent developments.} 

\FullConference{Corfu Summer Institute 2018 "School and Workshops on Elementary Particle Physics and Gravity"\\
		(CORFU2018)\\
		31 August - 28 September, 2018\\
		Corfu, Greece}

\begin{document}

\section{Introduction and discussion}

The spontaneous breaking of supersymmetry is one of the most challenging aspects of realistic model building within String Theory and Supergravity. 
The phenomenon itself, 
in four-dimensional N=1 supergravity, 
is well-understood and there are various models that can describe the supersymmetry breaking sector \cite{FVP,Kallosh:2014oja}. 
For example, 
if we study a single chiral superfield coupled to supergravity with K\"ahler potential and superpotential \cite{Polonyi:1977pj}\footnote{We will not review four-dimensional N=1 supergravity here, 
rather we refer the reader to the book of Wess and Bagger \cite{WB} for a superspace description of off-shell supergravity and the book of Freedman and Van Proeyen \cite{FVP} 
for a tensor calculus description. In this work we set $M_P=1$.} 
\be
K = \Phi \overline \Phi \, , \quad W = \mu \left( \Phi -\beta \right) \, , 
\ee
the system will have a stable Minkowski vacuum 
for $\beta = 2 - \sqrt 3$ 
with gravitino mass 
\be
m_{3/2} = \mu e^{\beta} \, .  
\ee
Because we are in a Minkowski vacuum the gravitino mass will also match with the supersymmetry breaking scale 
\be
{\rm F}_{\text{SUSY}} = \sqrt{ \langle {\cal V} \rangle + 3 m_{3/2}^2 } = \sqrt 3 m_{3/2} \, . 
\ee
For further properties of this system, which is referred to as the {\it Polonyi model}, see \cite{Polonyi:1977pj,FVP}. 
The Polonyi model is flexible enough, 
such that it allows for two types of important deformations. 
Firstly, 
one can slightly reduce the $\beta$ parameter in the superpotential and construct a stable de Sitter vacuum (see e.g. \cite{Akrami:2018ylq}). 
A second deformation is to include a $-\lambda |\Phi|^4$ term in the K\"ahler potential 
that will induce an extra contribution to the mass of the scalar proportional to $\lambda \mu^2$. 
The Polonyi model can be considered as an F-type supersymmetry breaking model in supergravity. 
This is because if we use the chiral superfield expansion 
\be
\Phi = A + \sqrt 2 \Theta \chi + \Theta^2 F \, , 
\ee
one can check that it is the auxiliary field $F$ that gets a vev and triggers the spontaneous supersymmetry breaking. 
In contrast, 
when the auxiliary field $M$ (or $F^0$ in the superconformal setup) of the supergravity multiplet gets a vev supersymmetry is not essentially broken, 
rather we get the anti de Sitter supergravity \cite{FVP}. 
Other notable type of F-term supersymmetry breaking models are the {\it no-scale} models \cite{Cremmer:1983bf}, 
and the models related to {\it higher order} F-potentials (see e.g. \cite{Koehn:2012ar,Farakos:2012qu,Nitta:2018yzb}).

Another important class of supersymmetry breaking models is the D-term type, 
which will also be the main topic of this contribution. 
Let us first remind the reader that an abelian vector multiplet coupled to supergravity is described by the 
real superfield $V$ that transforms under gauge transformations as 
\be
\label{gaugeV}
V \to V + i S - i \overline S \, , 
\ee
where the superfield $S$ is chiral. 
The vector superfield $V$ 
has a chiral superfield field strength given by 
\begin{equation}
{\cal W}_\alpha(V) = - \frac14 \left( \overline{\cal D}^2 - 8 {\cal R} \right) {\cal D}_\alpha V \, . 
\end{equation}
Here ${\cal R}$ is a chiral superfield of the supergravity sector with lowest component given by the auxiliary field $M$, 
that is ${\cal R}|=-M/6$. 
The component fields of the vector multiplet are then given by 
\begin{equation} 
\label{Wcomp}
{\cal W}_\alpha | = - i \lambda_\alpha \, ,  \quad 
{\cal D}_{(\alpha} {\cal W}_{\beta)}| = - 2 i \sigma^{ab\ \rho}_{\ \ \alpha} \epsilon_{\rho \beta} \hat D_a v_b \, , 
\quad 
{\cal D W}| = - 2 {\rm D} \, , 
\end{equation}
where $\hat D_b v_a$ is  the supercovariant derivative of the abelian vector and can be found in \cite{WB}. 
The Weyl spinor $\lambda$ is the gaugino and D is a real scalar auxiliary field. 
Under a supersymmetry transformation the gaugino transforms as 
\be
\delta \lambda_\alpha = -2 i \epsilon_\alpha \text{D} - 2 (\sigma^{ab} \epsilon)_\alpha \hat F_{ab} \, , 
\ee
where $\hat F_{ab} = \hat D_a v_b - \hat D_b v_a$. 
When supersymmetry is broken by the D-term then the gaugino will contribute to the goldstino, 
and we can always set the former to vanish by a gauge transformation. 
This gauge choice is identified with the unitary gauge when the supersymmetry breaking is sourced completely by the D-term, 
therefore in such case we have: 
\be
\label{UUGG}
\text{Unitary gauge for pure D-term breaking:} \quad \quad \lambda_\alpha = 0 \, . 
\ee

The simplest construction that allows to introduce a D-term supersymmetry breaking is to embed the 
Fayet--Iliopoulos (FI) model \cite{Fayet:1974jb} in supergravity. 
The central ingredient of the FI model is the FI term, 
which in global supersymmetry is given by 
\be
{\cal L}_{FI} = -2 \sqrt 2 \xi \int d^4 \theta V = - \sqrt 2 \xi \text{D} \, . 
\ee
In practice one simply has to embed a term of the form $e \text{D}$  in supergravity, 
where $e$ is the determinant of the vielbein. 
Because of the properties of the real superspace density $E$ of the old-minimal supergravity, 
a term of the form 
\be
\label{wbFI}
\int d^4 \theta \, E \, V  \, , 
\ee
is not gauge invariant. 
Indeed because for the old-minimal superspace formulation of supergravity we have 
\be
\int d^4 x \, d^4 \theta \, E \, S \ne 0 \, ,  
\ee 
the term \eqref{wbFI} is not invariant under \eqref{gaugeV}.

The embedding of the FI term in old-minimal N=1 supergravity is however possible if we gauge the $U(1)_{\rm R}$ symmetry. 
This was originally done by Freedman in \cite{Freedman:1976uk}. 
Let us first recall that under a super-Weyl rescaling the superspace integral together with the density transform as \cite{WB} 
\be
\int d^4 \theta \, E \, \to \, \int d^4 \theta \, E \, \text{e}^{2 \Sigma + 2 \overline \Sigma} \, . 
\ee
As a result the term \cite{Stelle:1978wj,VanProeyen:1979ks} 
\be
{\cal L}_{\text{standard FI}} = -3  \int d^4 \theta \, E \, \text{e}^{2 \sqrt 2 \xi V/3} 
= -3  \int d^4 \theta \, E 
-2 \sqrt 2 \xi \int d^4 \theta \, E \, V + {\cal O}(V^2)   \, , 
\ee
is gauge invariant if we have 
\be
\label{R-gauging}
\frac{2 \sqrt 2}{3} iS \xi = -2 \Sigma \, . 
\ee
In this way we arrive at the {\it Freedman model} \cite{Freedman:1976uk} that is given in superspace by 
\be
{\cal L} = -3  \int d^4 \theta \, E \, \text{e}^{2 \sqrt 2 \xi V/3} 
+ \frac14 \left( \int d^2 \Theta \, 2{\cal E} \, {\cal W}^2 + c.c. \right) \, . 
\ee 
Notice that the superspace kinetic term of the vector multiplet is gauge invariant and super-Weyl invariant independently, 
because under a super-Weyl transformation we have 
\be
\int d^2 \Theta \, 2{\cal E} \, \to \int d^2 \Theta \, 2{\cal E} \, \text{e}^{6 \Sigma} \, , \quad {\cal W}_\alpha(V) \, \to \,  {\cal W}_\alpha(V) \, \text{e}^{-3 \Sigma} \, . 
\ee
In component form we have 
\be
\label{freedman}
\begin{aligned}
e^{-1} {\cal L}\Big{|}_{\lambda=0} 
= &   -\frac12 R 
+ \frac12 \epsilon^{klmn} \left( \overline \psi_k \overline \sigma_l D_m \psi_n 
- \psi_k \sigma_l D_m \overline \psi_n  \right)  
\\
&  -\frac14 F_{mn} F^{mn} +  \frac{i}{2} e \xi \epsilon^{klmn} \overline \psi_k \overline \sigma_l \psi_n v_m -  \xi^2  \, , 
\end{aligned}
\ee
where $D_m$ is the covariant derivative for Lorentz indices that includes the spin-connection $\omega_{ma}^{\ \ \ b}(e,\psi)$. 
An inspection of the Lagrangian \eqref{freedman} shows that the {\it R-symmetry is gauged}. 
This happens because on one hand the gravitino is by definition charged under the R-symmetry \cite{WB,FVP}, 
while on the other hand we see that the Lagrangian  \eqref{freedman} contains a {\it minimal coupling} 
between the gravitino and the FI abelian gauge vector $v_m$. 
Therefore, 
for the self-consistency of this coupling, 
the gauging of the R-symmetry by the FI abelian gauge vector takes place.\footnote{This is of course exactly what the identification \eqref{R-gauging} implies.} 
For a discussion on the R-symmetry gauging and the FI terms see \cite{Barbieri:1982ac,VanProeyen:2004xt}. 
The supersymmetry breaking scale in the Freedman model is 
\be
{\rm F}_{\text{SUSY}} = \sqrt{ \langle {\cal V} \rangle + 3 m_{3/2}^2 } = \langle {\cal V} \rangle = \xi^2 \, . 
\ee
Notice that the gravitino has no mass term in \eqref{freedman}.

In the next two sections of this contribution we will review the properties of a new construction  
that allows to embed the FI term in N=1 supergravity without gauging the R-symmetry.

\section{New FI terms in N=1}

To explain the rationale behind the construction of the new FI term presented in \cite{Cribiori:2017laj}, 
we will start by discussing the {\it non-linear realizations} of supersymmetry within supergravity \cite{Samuel:1982uh} 
\footnote{We will not review here the non-linear realizations of supersymmetry, 
however, for a recent review see \cite{Cribiori:2019cgz}.}. 
When we have an N=1 supergravity theory where supersymmetry is spontaneously broken we can 
define a spinor superfield $\Gamma_\alpha$ 
with the properties 
\be
\label{SW}
\begin{aligned}
{\cal D}_\alpha \Gamma_\beta  &= \epsilon_{\beta \alpha} \left( 1 - 2 \, \Gamma^2 {\cal R} \right)  \, , 
\\
\overline{\cal D}^{\dot \beta} \Gamma^\alpha &= 2 i \, \left( \overline \sigma^a \, \Gamma \right)^{\dot \beta} \, {\cal D}_a \Gamma^\alpha 
+ \frac12 \, \Gamma^2 {\cal G}^{\dot \beta \alpha} \, .  
\end{aligned}
\ee
Here ${\cal G}_a$ is a superfield of the supergravity sector, 
which has lowest component ${\cal G}_a| = -b_a/3$. 
The important properties of this superfield are that the only independent component field is described by the lowest component 
\be
\gamma_\alpha = \Gamma_\alpha | \, , 
\ee
and that the supersymmetry transformation of the lowest component have the form 
\be
\delta \gamma_ \alpha = \epsilon_\alpha(x) + \dots 
\ee
Indeed, 
the constraints \eqref{SW} imply that the descendant component fields of the $\Gamma_\alpha$ 
superfield are composite fields that depend only on $\gamma_\alpha$ and on the supergravity sector (see e.g. \cite{Samuel:1982uh,DallAgata:2016syy,Farakos:2017bxs}). 
The simplest construction with the $\Gamma$ superfield is given by 
\be
\label{PPP}
\int d^4 \theta \, E \, \Gamma^2 \overline \Gamma^2 = e + {\cal O}(\gamma, \overline \gamma) \, ,  
\ee
thus gives a positive contribution to the vacuum energy once it is coupled to 
supergravity \cite{Samuel:1982uh}. 
In addition, 
if the unitary gauge is $\gamma=0$, then the term \eqref{PPP} will only contribute to the cosmological constant, as the $\gamma$ terms will drop out. 
If however we have a superfield ${\cal U}$ inside the superspace integral together with the $\Gamma$, 
then we can generate terms with the lowest component of ${\cal U}|=U$ to be the only non-vanishing 
contribution in the $\gamma=0$ gauge. 
That is we have 
\be
\label{EU}
\int d^4 \theta \, E \, {\cal U} \, \Gamma^2 \overline \Gamma^2 = e \, U + {\cal O}(\gamma, \overline \gamma) \, . 
\ee
The rationale behind the construction of the new FI term presented in \cite{Cribiori:2017laj} can now be explained. 
{\it First}, 
one has to insert a specific superfield in \eqref{EU} with the lowest component given by D, 
and {\it second}, one has to relate the $\gamma$ fermions with the gaugini.

Let us first relate the $\gamma$ fermions to the gaugini. 
We can do this directly in superspace by simply postulating 
\be
\label{GW2}
\Gamma_\alpha = - 2 \frac{{\cal D}_\alpha {\cal W}^2}{{\cal D}^2 {\cal W}^2} \, . 
\ee
One can check that \eqref{GW2} does indeed satisfy the constraints \eqref{SW}. 
For the lowest component of $\Gamma$ we have 
\be
\gamma_\alpha = \frac{2i {\cal Z}_{\alpha \beta}}{{\cal Z}^{\rho \sigma} {\cal Z}_{\rho \sigma}} \, \lambda^\beta + \text{3-fermi terms} \, , 
\ee 
where 
\be
{\cal Z}_{\alpha \beta} =  {\cal D}_\beta {\cal W}_\alpha \, . 
\ee
From \eqref{Wcomp} we see that 
\be
\label{L/D}
\gamma_\alpha = \frac{i \lambda_\alpha}{2 {\text{D}}} + \dots 
\ee
From equation \eqref{L/D} we see that if the vev of D is vanishing then we cannot use the form of $\Gamma$ given by \eqref{GW2}. 
If on the contrary we insist on having a consistent effective field theory while we will always use \eqref{GW2}, 
then we will require 
\be
\label{DNV}
\langle {\text D} \rangle \ne 0 \, . 
\ee
As we will shortly see, 
the self-consistency of the full construction of \cite{Cribiori:2017laj} will guarantee \eqref{DNV}.

We now turn to the second step of the construction of the new FI term. 
We want to have a superfield with lowest component given by the auxiliary field D. 
This is easily achieved because from \eqref{Wcomp} we see that the superfield we are 
interested in is given by ${\cal D}^\alpha{\cal W}_\alpha$.  
We therefore have 
\be
\label{EU}
\int d^4 \theta \, E \, {\cal DW} \, \Gamma^2 \overline \Gamma^2 = -2 \,e \, {\text D} + {\cal O}(\gamma, \overline \gamma) \, , 
\ee
or in the way it is constructed in \cite{Cribiori:2017laj} we will have 
\be
\label{newFI}
{\cal L}_{\text{new FI}} =  8 \sqrt 2 \xi  \int d^4\theta \, E \, \frac{{\cal W}^2\overline{{\cal W}}^2}{{\cal D}^2 {\cal W}^2 \overline{\cal D}^2 \overline{\cal W}^2} {\cal D}W \, . 
\ee
One can go from \eqref{EU} to \eqref{newFI} by taking into account 
the algebraic identity 
\be
\Gamma^2 \overline \Gamma^2 \equiv 16 \frac{
{\cal W}^2 \overline {\cal W}^2 
}{{\cal D}^2 {\cal W}^2 \overline {\cal D}^2 \overline {\cal W}^2} \, ,
\ee 
that is derived by using \eqref{GW2}. 
Clearly once we expand \eqref{newFI} in components we will have 
\be
\label{LFI}
{\cal L}_{\text{new FI}} =  - \sqrt 2 \xi  \,e \, {\text D} + {\cal O}(\lambda, \overline \lambda) \, . 
\ee
It is important to stress that the fermionic terms in \eqref{LFI}  are in general divided by the auxiliary field D.  
However, 
as the kinetic term for the vector multiplet contains a term quadratic in the auxiliary field D, 
that is 
\be
{\cal L}_{\text{D}} =  e \frac12 \text{D}^2  - \sqrt 2 e \xi \, {\text D} \, , 
\ee 
then by integrating out D we will generically find it has a non-vanishing vev 
\be
{\text{D}} = \sqrt 2  \xi \, , 
\ee 
and therefore the full construction will be self-consistent.

We can now consider the simplest supergravity model with the new FI term. 
The superspace Lagrangian is given by 
\be
\begin{aligned}
{\cal L} =&  - 3 \left( \int d^2 \Theta \, 2 {\cal E} \, {\cal R} + c.c. \right)  
+  \left( \int d^2 \Theta \, 2 {\cal E} \, W_0 + c.c. \right)  
\\
&   + \frac14 \left( \int d^2 \Theta \, 2 {\cal E} \, {\cal W}^2(V) + c.c. \right) 
+ 8 \sqrt 2 \xi  \int d^4\theta \, E \, \frac{{\cal W}^2\overline{{\cal W}}^2}{{\cal D}^2 {\cal W}^2 \overline{\cal D}^2 \overline{\cal W}^2} {\cal D}{\cal W} \, . 
\end{aligned}
\ee
Once we write the theory in component form and integrate out the auxiliary fields we find 
\be
\label{NEWDTERM}
\begin{aligned}
e^{-1} {\cal L} \Big{|}_{\lambda=0}  
=&   -\frac12 R 
+ \frac12 \epsilon^{klmn} \left( \overline \psi_k \overline \sigma_l D_m \psi_n 
- \psi_k \sigma_l D_m \overline \psi_n  \right)  
\\
&   -\frac14 F_{mn} F^{mn} - \left( \xi^2 -3 |W_0|^2 \right) 
- \overline W_0 \psi_a \sigma^{ab} \psi_b - W_0 \overline \psi_a \overline \sigma^{ab} \overline \psi_b  \, . 
\end{aligned}
\ee
We can highlight some properties of the Lagrangian \eqref{NEWDTERM}: 
\begin{itemize}

\item No R-symmetry gauging.  

\item Arbitrary gravitino {\it Majorana} mass: 
\be
m_{3/2} = W_0 \, . 
\ee

\item The cosmological constant can be tuned: 
\be 
{\cal V} = \xi^2 -3 |m_{3/2}|^2 \, . 
\ee

\item Independent vacuum energy from supersymmetry breaking scale: 
\be
{\rm F}_{\text{SUSY}} = \sqrt{ \langle {\cal V} \rangle + 3 m_{3/2}^2 } = \xi^2 \, . 
\ee

\item No {\it smooth} supersymmetric limit.

\item Electric-magnetic duality: 
\be
F_{mn} \rightarrow \epsilon_{mnkl} F^{kl} \, . 
\ee

\end{itemize}

This concludes the simplest model that we can construct with the use of the new Fayet--Iliopoulos term of \cite{Cribiori:2017laj}. 
Developments related to cosmology can be found in 
\cite{Aldabergenov:2017hvp,Antoniadis:2018cpq,Antoniadis:2018oeh,Abe:2018rnu,Antoniadis:2018vtd,Ishikawa:2019pnb}, 
further progress in the study of matter couplings can be found in \cite{Aldabergenov:2018nzd,Abe:2018plc}, 
and variations of the construction can be found in \cite{Kuzenko:2017zla,Kuzenko:2018jlz,Farakos:2018sgq,Kuzenko:2019vaw}. 
Let us note that these constructions are only known for four-dimensional N=1 supergravity.

\section{Anti D3-brane interpretation}

In this section we will study the new FI term when matter fields are included and we will also study the new FI term in the global limit. 
These two aspects of the new FI term hint towards an interpretation in terms of anti D3-branes.

\subsection{Matter coupling and scalar potential}

We now introduce a single chiral superfield $\Phi$ in the theory and we will study a Lagrangian of the form 
\be
\label{mmm}
\begin{aligned}
{\cal L} =&  - 3 \int d^4 \theta \, E \, {\text e}^{-K/3} 
+  \left( \int d^2 \Theta \, 2 {\cal E} \, W(\Phi) + c.c. \right)  
\\
&   + \frac14 \left( \int d^2 \Theta \, 2 {\cal E} \, {\cal W}^2(V) + c.c. \right) 
+ 8 \sqrt 2 \xi  \int d^4\theta \, E \, \frac{{\cal W}^2\overline{{\cal W}}^2}{{\cal D}^2 {\cal W}^2 \overline{\cal D}^2 \overline{\cal W}^2} {\cal D}W \, . 
\end{aligned}
\ee
We should point out that even though $K$ appears as the K\"ahler potential in \eqref{mmm}, 
the K\"ahler invariance is explicitly broken by the new FI term. 
This however does not mean the theory is not supersymmetric. 
It is also possible to restore the K\"ahler invariance by introducing appropriate ${\text e}^K$ factors as has been pointed out in \cite{Antoniadis:2018cpq}. 
Here we will review the matter couplings as presented in \cite{Cribiori:2017laj} which are described by the Lagrangian \eqref{mmm}. 
When we reduce \eqref{mmm} to component form we find {\it in the unitary gauge} \eqref{UUGG} that 
the Lagrangian has exactly the same form as in standard supergravity 
and the only difference appears in the scalar potential that is given by 
\be
\label{newV}
{\cal V} = {\cal V}_{\substack{\text{Standard} \\\text{SUGRA}}} + \xi^2 \, {\text e}^{2K/3}\, . 
\ee
The very interesting property of the new term that enters the scalar potential \eqref{newV} is that it 
gives rise to an {\it uplift}, 
and as we will see now it matches with the uplift that is ascribed to an anti D3-brane in the KKLT scenario \cite{Kachru:2003aw,Kachru:2003sx}. 
Indeed, 
if we assume a no-scale K\"ahler potential \cite{Cremmer:1983bf} (as in KKLT), 
that is 
\be
K = -3 \ln(\Phi + \overline \Phi) \, , 
\ee
then the uplift term gives to the scalar potential the form 
\be
\label{antiD3}
{\cal V} = {\cal V}_{\substack{\text{Standard} \\\text{SUGRA}}} +  \frac{\xi^2}{(A + \overline A )^2} \, . 
\ee
Note that the scalar potential \eqref{antiD3} is usually constructed by introducing non-linear realizations, 
either for example within the {\it constrained superfields} setup \cite{Ferrara:2014kva,Kallosh:2014wsa,Bergshoeff:2015jxa}, 
or in the {\it Goldstino brane} setup \cite{Bandos:2016xyu}. 
In both cases the effective supergravity theory is expected to capture the impact of the anti D3-brane in the KKLT scenario. 
However, 
the construction of a scalar potential \eqref{antiD3} with the new FI term of \cite{Cribiori:2017laj} 
has been achieved without explicitly invoking a non-linear realization of supersymmetry 
as the uplift effect is induced by a standard N=1 abelian vector multiplet and the field content of the Lagrangian \eqref{mmm} is supersymmetric.

Let us note that in \cite{DallAgata:2013jtw} a new kind of no-scale models can be introduced where the K\"ahler potential 
has the form $K = -2 \ln(\Phi + \overline \Phi)$ and there is a gauging of the isometry $\Phi \to \Phi + iq$, 
with real $q$, 
by the abelian vector multiplet $V$. 
Alternatively, the new FI term can be also utilized, 
and no-scale models can be constructed when \eqref{newFI} is included in a setup with $K = - \ln(\Phi + \overline \Phi)$ 
and with no gauging introduced \cite{Aldabergenov:2019hvl}.

\subsection{Super Born--Infeld and alternative Bagger--Galperin}

In this subsection we will review the results of \cite{Cribiori:2018dlc} where it was shown that when the new FI term \eqref{newFI} 
is studied in the global limit then it can be identified as the source of supersymmetry breaking in the Bagger--Galperin (BG) model \cite{Bagger:1996wp}. 
The Bagger--Galperin model is an N=1 supersymmetric theory constructed by an abelian vector multiplet with a 
bosonic sector matching the Born--Infeld. 
Generic supersymmetric theories that have a bosonic sector with this property were constructed in \cite{Cecotti:1986gb}, 
but the BG construction is a special class of these models that has a {\it second non-linearly realized supersymmetry} 
of the Volkov--Akulov (VA) type \cite{Volkov:1973ix}. 
This property of the BG action allows to identify it as the effective action of a space-filling (anti) D3-brane with truncated spectrum 
(see e.g. the discussions in \cite{Rocek:1997hi,Kallosh:2016aep}). 
Further developments related to the BG action can be found in 
\cite{Bandos:2001ku,Klein:2002vu,Antoniadis:2008uk,Kuzenko:2009ym,Ferrara:2014oka,Bellucci:2015qpa,Cribiori:2018jjh,Antoniadis:2019gbd}.

The BG model can be written in terms of a standard non-linear realization of supersymmetry (see e.g. \cite{Klein:2002vu} for a recent review) 
in the form \cite{Bellucci:2015qpa,Cribiori:2018dlc,Antoniadis:2019gbd} 
\begin{equation} 
\label{BG1}
S_\text{BG}=- \beta m \int d^4 x \det[A_m^a] \left( 
1+ 
\sqrt{-\det{\left(\eta_{ab}+\frac{1}{m}{\cal F}_{ab}\right) }} 
\right) \, , 
\end{equation} 
where $\beta$ and $m$ are real constants. 
The expressions appearing in \eqref{BG1} are 
\begin{equation}
\label{AcompAM}
A_m^a = \delta_m^a 
- i \partial_m\chi\sigma^a\overline\chi 
+ i \chi\sigma^a\partial_m\overline\chi \,  , 
\end{equation}
with the fermion $\chi$ describing the standard VA goldstino that transforms under supersymmetry as \cite{Volkov:1973ix} 
\be
\delta \chi_\alpha = \epsilon_\alpha 
- i \left( \chi \sigma^m \overline \epsilon - \epsilon \sigma^m \overline \chi \right) \partial_m \chi_\alpha \, . 
\ee
The field strength of the abelian vector is given by 
\begin{equation}
{\cal F}_{ab}=(A^{-1})^m_a(A^{-1})^n_b\left[\partial_{m} u_{n}-\partial_{n} u_{m}\right] \, , 
\end{equation}
where the $u_m$ transforms under supersymmetry as 
\be
\delta u_m=- i \left(\chi\sigma^n\overline\epsilon
-\epsilon\sigma^n\overline\chi\right)\partial_n u_m 
- i \partial_m\left(\chi \sigma^n\overline\epsilon 
-\epsilon \sigma^n\overline\chi\right)u_n \, . 
\ee
The form of the second supersymmetry that transforms the vector $u_m$ into the fermion $\chi_\alpha$ has been derived in \cite{Bellucci:2015qpa}. 
In \cite{Cribiori:2018dlc} an alternative formulation of the BG action was presented that has the form 
\begin{equation}
\begin{aligned} 
S_{\overline{\text{BG}}}=&\frac{\beta}{4m} \int d^4 x\, \left( d^2\theta\, {\cal W}^2 + c.c. \right) 
+ 16  \beta  \int d^4x\, d^4\theta\, \frac{{\cal W}^2\overline{\cal W}^2}{D^2{\cal W}^2\overline D^2\overline{\cal W}^2}D^\alpha {\cal W}_\alpha\\
& + 16 \beta m \int d^4x\, d^4\theta\, \frac{{\cal W}^2\overline{\cal W}^2}{D^2{\cal W}^2\overline D^2\overline {\cal W}^2}\left\{1+\frac{1}{4m^2} f_{ab} f^{ab} -\sqrt{-\det \left(\eta_{ab} + \frac{1}{m} f_{ab} \right) } \right\} \,,
\label{BG2}
\end{aligned}
\end{equation}
where we have defined the superfield 
\begin{equation}
f_{ab}=\frac{i}{4}\sigma_{ab\,\gamma}{}^\alpha\varepsilon^{\gamma\beta}\left(D_\alpha {\cal W}_\beta +D_\beta {\cal W}_\alpha\right)+ c.c. \,. 
\end{equation}
Notice that the first line of \eqref{BG2} has the same structure as the new D-term of \cite{Cribiori:2017laj}. 
When one writes \eqref{BG2} in component form and integrates out the auxiliary fields it will reduce to \eqref{BG1} after appropriate field redefinitions \cite{Cribiori:2017ngp}. 
This procedure has been performed in detail in \cite{Cribiori:2018dlc}. 
This finding further supports the interpretation of the new FI term as the source of supersymmetry breaking 
related to the effective theory of an (anti) D3-brane 
and also allows to identify the FI parameter in terms of the brane tension and the $\alpha'$ \cite{Cribiori:2018dlc}. 
In \cite{Cribiori:2018dlc} this Lagrangian is also embedded in four-dimensional N=1 supergravity.

\acknowledgments  

I would like to thank Niccol\`o Cribiori for discussions. 
This work is supported from the KU Leuven C1 grant ZKD1118 C16/16/005.

\end{document}